\documentclass[twocolumn,showpacs,showkeys,preprintnumbers]{revtex4}
\def\be{\begin{equation}}
\def\ee{\end{equation}}
\def\beq{\begin{eqnarray}}
\def\eeq{\end{eqnarray}}
\def\n{\nonumber}
\def\bay{\begin{array}}
\def\eay{\end{array}}

\begin{document}

\preprint{CIRI/02-sw03}
\title{A New General Relativistic Cosmology}

\author{Sanjay M. Wagh}% and Ravindra V Saraykar}
\affiliation{Central India Research Institute, Post Box 606,
Laxminagar, Nagpur 440 022, India\\
E-mail:cirinag@nagpur.dot.net.in }

\date{October 14, 2002}
\begin{abstract}
In this work, we outline a new general relativistic cosmology. In
this cosmology, the universe originates in the infinite past from
sparsely distributed neutral matter and ends in the infinite
future as a hot, relativistic plasma. The spatial distribution of
matter on the ``initial" hyper-surface is {\em arbitrary}. Hence,
observed structures can arise in this cosmology from suitable
``initial" density contrast. The red-shifts of different objects
in this cosmology are indicative of their different states of
collapse and need not possess any correlation to their distance
from the observer. Further, the microwave background radiation
arises in this cosmology as thermalized radiation from all the
radiating matter in the universe. This cosmology predicts that the
temperature of the microwave background increases with time. Thus,
any conclusive evidence that the temperature of the Microwave
Background Radiation was more in the past can falsify this
cosmology.
\end{abstract}

 \maketitle

\newpage
\section{Introduction} In the 3+1 formulation \cite{mtw} of
General Relativity,  {\em source\/} matter or energy data can be
specified on an ``initial" spacelike hyper-surface and, that data
can be evolved using the Einstein field equations. Different
four-dimensional spacetime geometries are obtainable for different
initial data.

When matter or energy is present over all of the initial
hyper-surface, we obtain a ``cosmological"  situation or a
``cosmological" spacetime.

A famous is the case of maximally symmetric, homogeneous and
isotropic, Friedmann-Lema\'\i tre-Robertson-Walker (FLRW)
spacetime of the Big Bang Cosmology (BBC) \cite{mtw, sussp,
stdbooks}.

Now, to fix ideas, recall that the Newtonian Law of Gravitation is
a statement of the force of attraction between two mass points in
space. The presence of other mass-points does not alter the
statement of this Law of Gravitation. Moreover, re-distribution of
mass-points does not affect the statement of this Law of
Gravitation.

General Relativity prescribes a spacetime geometry for a Law of
Gravitation. In analogy with the Newtonian case, we therefore seek
a spacetime that ``does not change" its geometrical or physical
characters when matter or energy is differently distributed on its
initial hyper-surface. Moreover, the requirement that the addition
of ``more" matter or energy does not change the geometrical
characters of this spacetime means that it is also a
``cosmological" spacetime.

But, we note that the FLRW spacetime does {\em change\/}  its
geometrical characters when matter is differently distributed,
say, in some inhomogeneous way, on its initial hyper-surface.

Therefore, we consider here a Petrov-type D, cosmological
spacetime which ``does not change" its geometrical characters
under redistribution of matter on its initial hyper-surface and
explore the cosmology implied by it.

\section{Spacetime metric} \label{spacetime} Consider the
spacetime metric: \beq ds^2 &=& -X^2Y^2Z^2dt^2
+\gamma^2{X'}^2Y^2Z^2B^2 dx^2 \n
\\&\phantom{+}&\qquad +\;\gamma^2X^2\bar{Y}^2 Z^2C^2dy^2\n
\\&\phantom{+}& \qquad\quad +\;\gamma^2X^2Y^2\tilde{Z}^2D^2\,dx^2 \label{genhsp}
\eeq where $X\equiv X(x)$, $Y\equiv Y(y)$, $Z\equiv Z(z)$,
$B\equiv B(t)$, $C\equiv C(t)$, $D\equiv D(t)$ and $\gamma$ is a
constant. Also, $X'=dX/dx$, $\bar{Y}=dY/dy$ and $\tilde{Z}=dZ/dz$.

\subsection*{Singularities and Degeneracies of (\ref{genhsp})}
There are, in general, two types of singularities of the metric
(\ref{genhsp}). The first type is when any of the temporal
functions $B$, $C$, $D$ is vanishing and the second type is when
any one of the spatial functions $X$, $Y$, $Z$ is vanishing.
Singularities of the second type constitute a part of the initial
data, singular initial data, for (\ref{genhsp}). On the other
hand, vanishing of the temporal functions is a singular
hyper-surface of (\ref{genhsp}).

The locations for which the spatial derivatives vanish are,
however, coordinate singularities. The curvature invariants of
(\ref{genhsp}) do not blow up at such locations. Therefore, before
such coordinate singularities are reached, we may transform
coordinates to other suitable ones.

There are also obvious degenerate metric situations when any of
the spatial functions is infinite for some range of the
coordinates.

{\em In what follows, we shall assume, unless stated explicitly,
that there are no singular initial-data and that there are no
degenerate situations for the metric (\ref{genhsp}).}

The metric (\ref{genhsp}) admits three hyper-surface orthogonal
spatial homothetic Killing vectors (HKVs) \beq {\bf {\cal X}} &=& (0, \frac{X}{\gamma X'}, 0, 0) \label{genhkv1} \\
{\bf {\cal Y}} &=& (0, 0, -\frac{Y}{\gamma \bar{Y}}, 0) \label{genhkv2} \\
{\bf {\cal Z}} &=& (0, 0, 0, -\frac{Z}{\gamma\tilde{Z}})
\label{genhkv3} \eeq

Now, the existence of three HKVs is equivalent to two Killing
vectors (KVs) and one HKV. Then, the metric (\ref{genhsp}) can be
expressed in a form that displays the existence of KVs explicitly.
However, we will use the form (\ref{genhsp}) in this paper.

The spacetime of (\ref{genhsp}) is required, by definition, to be
locally flat at all of its points. In general, this will require
some conditions on the metric functions $X$, $Y$, $Z$.

Then, the Einstein tensor has the components \beq G_{tt}&=&
-\frac{1}{\gamma^2B^2}-\frac{1}{\gamma^2C^2}
-\frac{1}{\gamma^2D^2} \n \\ &\phantom{m}& \qquad +
\frac{\dot{B}\dot{C}}{BC} +
\frac{\dot{B}\dot{D}}{BD} +\frac{\dot{C}\dot{D}}{CD} \label{gtt}\\
G_{xx}&=&\frac{\gamma^2B^2X'^2}{X^2}
\left[-\frac{\ddot{C}}{C}-\frac{\ddot{D}}{D} -
\frac{\dot{C}\dot{D}}{CD} \right. \n \\ &\phantom{+}& \qquad
\left. +\frac{3}{\gamma^2B^2}+ \frac{1}{\gamma^2C^2} +
\frac{1}{\gamma^2D^2}\right] \label{gxx}
\\G_{yy}&=& \frac{\gamma^2C^2\bar{Y}^2}{Y^2}
\left[-\frac{\ddot{B}}{B}-\frac{\ddot{D}}{D} -
\frac{\dot{B}\dot{D}}{BD}\right. \n \\ &\phantom{+}& \qquad \left.
+\frac{3}{\gamma^2C^2}+ \frac{1}{\gamma^2B^2} +
\frac{1}{\gamma^2D^2}\right] \label{gyy}
\\G_{zz}&=& \frac{\gamma^2D^2\tilde{Z}^2}{Z^2}
\left[-\frac{\ddot{B}}{B}-\frac{\ddot{C}}{C} -
\frac{\dot{B}\dot{C}}{BC} \right. \n \\ &\phantom{+}& \qquad
\left. +\frac{3}{\gamma^2D^2}+
\frac{1}{\gamma^2B^2} + \frac{1}{\gamma^2C^2}\right] \label{gzz}\\
G_{tx}&=&2\frac{\dot{B}X'}{BX} \qquad\qquad
G_{ty}=2\frac{\dot{C}\bar{Y}}{CY}\label{fluxes1} \\
G_{tz}&=&2\frac{\dot{D}\tilde{Z}}{DZ} \label{fluxes2}\\
G_{xy}&=&2\frac{X'\bar{Y}}{XY} \qquad\qquad
G_{xz}=2\frac{X'\tilde{Z}}{XZ} \label{stresses1} \\
G_{yz}&=&2\frac{\bar{Y}\tilde{Z}}{YZ} \label{stresses2} \eeq where
an overhead dot is used to denote a time-derivative.

The four-velocity of the co-moving observer is $U^a =
{\delta^a}_t/XYZ$. The expansion for (\ref{genhsp}) is: $\Theta
\propto \dot{B}/B + \dot{C}/C + \dot{D}/D$.

Now, let the energy-momentum tensor be: \beq
T_{ab} &=& (\rho + p+ {\cal B})U_aU_b + (p+{\cal B})\, g_{ab}\n \\
&\phantom{=}&\qquad + q_aU_b + q_bU_a + \Pi_{ab} \eeq where $\rho$
is the energy density, $p$ is the isotropic pressure, $q_a = (0,
q_x, q_y, q_z)$ is the energy-flux four-vector, ${\cal B}$ is the
bulk-viscous pressure, $\Pi_{ab}$ is the anisotropic stress tensor
etc. Note also that $q_aU^a = 0$ and that $\Pi_{ab}U^b = {\Pi^a}_a
= 0$. Thus, $\Pi_{ab}$ is symmetric, spatial and trace-free.

Now, from the Einstein field equations, with $8\pi G/c^4=1$, it
follows that $\rho = G_{tt}/X^2Y^2Z^2$. The isotropic pressure is
obtainable from the combination of (\ref{gxx}), (\ref{gyy}) and
(\ref{gzz}). The energy-fluxes $q_x$, $q_y$ and $q_z$ are related
to corresponding components of the Einstein tensor in
(\ref{fluxes1})-(\ref{fluxes2}) and the anisotropic stresses are
related to corresponding components of the Einstein tensor in
(\ref{stresses1})-(\ref{stresses2}). All the spatial functions
$X$, $Y$, $Z$ are {\em not\/} determined by the field equations
and, hence, are {\em arbitrary}.

Then, the ``initial hyper-surface" is given by
$\dot{B}=\dot{C}=\dot{D} = 0$. Now, the energy-fluxes are
vanishing for the initial hyper-surface. The anisotropic stresses
get determined from the spatial functions $X$, $Y$ and $Z$ as a
part of the initial data.

Now, to be able to explicitly solve the corresponding ordinary
differential equations, we require {\em physical\/} information
about matter or energy. To provide for the required information of
``physical" nature is a non-trivial task in general relativity
just as it is for the Newtonian gravity \cite{stars}. The details
of these considerations are, of course, beyond the scope of the
present paper.

Also, the non-vanishing components of the Weyl tensor for
(\ref{genhsp}) are: \beq C_{txtx}
&=& \frac{B^2\gamma^2X'^2Y^2Z^2}{6}\,F(t) \label{weyl1} \\
C_{tyty} &=& \frac{C^2\gamma^2X^2\bar{Y}^2Z^2}{6}\,G(t) \label{weyl2} \\
C_{tztz} &=& \frac{D^2\gamma^2X^2Y^2\tilde{Z}^2}{6}\,H(t)
\label{weyl3}
\\ C_{xyxy} &=&
-\,\frac{B^2C^2\gamma^4X'^2\bar{Y}^2Z^2}{6}\,H(t)\label{weyl4} \\
C_{xzxz} &=& -\,\frac{B^2D^2\gamma^4X'^2Y^2\tilde{Z}^2}{6}\,G(t)
\label{weyl5}
\\ C_{yzyz} &=&-\,
\frac{C^2D^2\gamma^4X^2\bar{Y}^2\tilde{Z}^2}{6}\,F(t)
\label{weyl6} \eeq where \beq F(t) &=& \frac{\ddot{C}}{C}
+\frac{\ddot{D}}{D}
-2\frac{\ddot{B}}{B} - 2\frac{\dot{C}\dot{D}}{CD} \n \\
&\phantom{m}& \qquad\quad\quad
+\frac{\dot{B}\dot{C}}{BC} + \frac{\dot{B}\dot{D}}{BD}  \\
G(t) &=& \frac{\ddot{D}}{D} +\frac{\ddot{B}}{B}
-2\frac{\ddot{C}}{C} + \frac{\dot{C}\dot{D}}{CD}\n \\
&\phantom{m}& \qquad\quad\quad
+\frac{\dot{B}\dot{C}}{BC} -2 \frac{\dot{B}\dot{D}}{BD} \\
H(t)&=& \frac{\ddot{B}}{B} +\frac{\ddot{C}}{C}
-2\frac{\ddot{D}}{D} +\frac{\dot{C}\dot{D}}{CD} \n \\
&\phantom{m}& \qquad\quad\quad +\frac{\dot{B}\dot{D}}{BD} -2
\frac{\dot{B}\dot{C}}{BC} \eeq

Now, for (\ref{genhsp}), it can be verified that the non-vanishing
Newman-Penrose (NP) complex scalars \cite{mtbh} are $\Psi_0 =
\Psi_4$ and $\Psi_2$.

Note that $\Psi_4 \neq 0$. Now, consider a NP-tetrad rotation of
Class II, with complex parameter $b$, that leaves the NP-vector
${\bf n}$ unchanged, and demand that the new value,
$\Psi_0^{(1)}$, of $\Psi_0$ is zero. Then, we have: \be\Psi_4\,b^4
+ 6 \Psi_2\,b^2 + \Psi_4 = 0 \ee This equation has {\em two\/}
distinct double-roots. Thus, the spacetime of (\ref{genhsp}) is of
Petrov-type D.

It is well-known that Penrose \cite{penroseweyl} is led to the
Weyl hypothesis on the basis of thermodynamical considerations, in
particular, those related to the thermodynamic arrow of time. On
the basis of these considerations, we may consider the Weyl tensor
to be ``some" sort of measure of the {\em entropy\/} in the
spacetime at any given epoch.

Now, note that the Weyl tensor, (\ref{weyl1}) - (\ref{weyl6}),
blows up at the singular hyper-surface of (\ref{genhsp}) but is
``vanishing" at the ``initial" hyper-surface for non-singular and
non-degenerate data since $\dot{B} = \dot{C} = \dot{D} = 0$ for
the ``initial" hyper-surface.

Then, we note that the {\em entropy\/} at the ``initial"
hyper-surface of (\ref{genhsp}) is ``zero" while that at its
singular hyper-surface is ``infinite".

This behavior of the Weyl tensor of (\ref{genhsp}) is in
conformity with Penrose's Weyl curvature hypothesis
\cite{penroseweyl}. Thus, the spacetime of (\ref{genhsp}) has the
``right" kind of thermodynamic arrow of time in it.

\subsection*{Other features of the metric (\ref{genhsp})}
The spatial distribution of matter is arbitrary for (\ref{genhsp})
and the evolution of non-singular and non-degenerate ``initial
data" is completely governed by only the temporal functions in
(\ref{genhsp}).

Therefore, changing matter distribution on the initial
hyper-surface does not change the geometrical properties of
(\ref{genhsp}) for non-singular and non-degenerate data. Thus,
(\ref{genhsp}) is the metric of the spacetime that we have been
looking for.

Now, non-gravitational processes are included, in general
relativity, via the energy-momentum tensor. Non-gravitational
processes primarily determine the relation of density and pressure
of matter - the equation of state. Matter properties, as are
applicable at any given stage of evolution of matter, such as an
equation of state of matter, the radiative characteristics of
matter etc.\ determine the temporal functions of (\ref{genhsp}).

For {\em physically realizable spacetime}, matter must pass
through different ``physical" stages of evolution, namely, from
dust to matter with pressure and radiation, to matter with
exothermic nuclear reactions etc. This is the causal and, hence,
physical, development of matter data. The spacetime of
(\ref{genhsp}) therefore describes physical development of matter
data.

Then, the initial hyper-surface, $t \to -\,\infty$, has sparsely
distributed matter for which the co-moving velocity of matter is
vanishing. The $x$, $y$, $z$ co-moving velocities are respectively
$\propto \dot{B},\,\dot{C},\,\dot{D}$ and, hence,
$\dot{B}=\dot{C}=\dot{D}=0$ on the initial hyper-surface of
(\ref{genhsp}).

Therefore, the temporal evolution in (\ref{genhsp}) for the
non-singular and non-degenerate data leads only to a temporal
singularity in the future.

Clearly, since the spatial distribution of matter is arbitrary for
(\ref{genhsp}), the present structures can evolve out of some
suitable density distribution.

The matter 4-velocity is \be
u^a\,=\,\frac{1}{XYZ\sqrt{\Delta}}\left(1, V^x, V^y, V^z\right)
\ee where $V^x = dx/dt$, $V^y = dy/dt$, $V^z = dz/dt$ and \beq
\Delta &=& 1 - \gamma^2\left[\frac{X'^2B^2{V^x}^2}{X^2}
+\frac{\bar{Y}^2C^2{V^y}^2}{Y^2} \right. \n \\ &\phantom{m}&
\left. \qquad\qquad\qquad\quad + \frac{\tilde{Z}^2D^2{V^z}^2}{Z^2}
\right] \label{delta} \eeq

Now, it can be inferred \cite{physical} that the velocity of
matter with respect to a co-moving observer is the speed of light
at the singular hyper-surface of (\ref{genhsp}). At the singular
hyper-surface of (\ref{genhsp}), $\Delta =1$. Then, a co-moving
observer also moves with the speed of light at the singular
hyper-surface.

After all, matter everywhere should become relativistic as
different mass condensates continue to grow (due to accretion onto
them) to influence the entire spacetime to become relativistic
everywhere. This is happening asymptotically for infinite
co-moving time.

Consequently, in the cosmology of (\ref{genhsp}), the universe
begins in the infinite past as ``cold" but ends as a soup of high
energy plasma and radiation in the infinite future. This is
evidently consistent with the behavior of the Weyl tensor as a
measure of the entropy.

Now, if $d\tau_{\scriptscriptstyle CM}$ is a small time duration
for a co-moving observer and if $d\tau_{\scriptscriptstyle RF}$ is
the corresponding time duration for the observer in the rest frame
of matter, then we have \be d\tau_{\scriptscriptstyle
CM}\;=\;\frac{d\tau_{\scriptscriptstyle RF}}{\sqrt{\Delta}}
\label{rshift1} \ee From (\ref{rshift1}), we also get the
red-shift formula \be \nu_{\scriptscriptstyle CM} =
\nu{\scriptscriptstyle RF}\,\sqrt{\Delta}\qquad z =
\frac{\nu{\scriptscriptstyle RF}}{\nu{\scriptscriptstyle
CM}}=\frac{1}{\sqrt{\Delta}}\label{rshift2} \ee in the spacetime
of (\ref{genhsp}) where $\nu_{\scriptscriptstyle CM}$ is the
frequency of a photon in the co-moving frame,
$\nu_{\scriptscriptstyle RF}$ is the frequency in the rest frame
and $z$ is the {\em total\/} red-shift of a photon.

Then, the red-shift, (\ref{rshift2}), depends on $X$, $Y$, $Z$
which determine the spatial density of matter and, of course, on
the temporal functions in (\ref{genhsp}). In the absence of
expansion in (\ref{genhsp}), the entire red-shift is indicative of
the state of collapse of matter at given co-moving time.

Now, we have sufficient characteristics of the spacetime of
(\ref{genhsp}) to discuss its cosmological implications to which
we now turn to.

\section{Cosmological implications}
Firstly, the universe of (\ref{genhsp}) need {\em not\/} be
expanding, rather, different matter condensates in it could simply
be contracting and collapsing onto more massive condensates.

This is true unless, of course, ``matter or energy" properties are
such as to lead to positive expansion, $\Theta$, for
(\ref{genhsp}). But, no matter condensate in (\ref{genhsp}) is
then collapsing in all dimensions. There must be, at least, one
direction in which matter must expand sufficiently rapidly for,
say, $\dot{B}/B$ to dominate over other negative terms in
$\Theta$.

Now, it has been aptly emphasized \cite{arp} that the red-shift of
an object should have an ``intrinsic" component, over and above
that due to expansion. Then, for (\ref{genhsp}), the red-shift,
(\ref{rshift2}), has both the contributions, from expansion, if
any, and from density characteristics of the object.

Note also that the entire red-shift is ``intrinsic" to the object
or is ``gravitational of origin" if the expansion of the universe
is non-existent for the spacetime of (\ref{genhsp}).

However, correlation of the red-shift with distance could also
arise from the initial density distribution if distant objects are
more collapsed in it. In this connection, we note that the
observational basis for the Hubble Law of expansion of the
universe has also been called into question from time to time
\cite{arp}.

For (\ref{genhsp}), two neighboring objects, physically connected
with luminous bridges, can be in different states of gravitational
collapse and, hence, can show different red-shifts. There is
therefore a natural explanation in the present cosmology for the
discordant or anomalous red-shifts of quasars and galaxies that
Arp \cite{arp} has been observing in quasar-galaxy associations.
The famous such case is that of NGC 7619 and its companion.

An important issue \cite{partridge, cobe} is that of the observed
Microwave Background Radiation (MBR). The early universe of
(\ref{genhsp}) is {\em cold\/} since matter is pressureless and
radiation-less close to the initial hyper-surface. Thus, the MBR
can only arise as thermalized radiation emitted by all the
radiating matter in the universe of (\ref{genhsp}).

Then, it is remarkable that a spacetime that does not change its
geometrical properties under redistribution of mass on its initial
hyper-surface leads to only such an explanation of the MBR. We
also note here that the energy density in the microwave,
$\rho_{mbr}$, is the same as that found in the starlight of our
own galaxy. This is suggestive \cite{arp} of the importance of
thermalization of light from radiating matter in the universe.

However, we note that all the earlier considerations of radiation
by warm dust producing the observed MBR consider an expanding
universe. As a result, these considerations face the problem of
energy supply needed to heat the dust for large red-shifts in an
expanding universe.

Another objection to the explanation of MBR as re-emission of
radiation from small-sized dust is that the spectrum of radiation
from such dust is unlikely to be black-body \cite{0209386} while
the spectrum of MBR is closely that of a black-body at $\sim 2.75
^o$K. Small dust particles are inefficient radiators at long
wavelengths.

However, sufficiently large ($\sim \,10-100\; \mu m$) sized dust
particles would thermalize the stellar radiation. But, for
expanding universe models, a prohibitively huge total dust mass is
obtained. That is, there is the need for large amounts of dust,
almost comparable to the density of heavy elements in the universe
\cite{partridge}. Such an era is not available for most
cosmological models.

Note, however, that the present cosmological model based on
(\ref{genhsp}) begins with an era of entire matter as being dust.
Consequently, for this model the above objections against the
explanation of MBR as thermalized emission by dust need to be
re-examined in details.

Now, since the entire matter in the present cosmology is expected
to attain the speed of light in the asymptotic future, we also
expect that the temperature of MBR should {\em increase\/} with
time. This is, therefore, a definite and falsifiable prediction of
this cosmology.

Any conclusive evidence that the temperature of MBR was more in
the past would also falsify this cosmology and, with it, the
premise that we need to seek a spacetime that ``does not change"
its geometrical characters when masses are differently distributed
on its initial hyper-surface.

As a separate remark, we note that, following the works of Ellis
and Sciama \cite{ellissciama}, there is an interpretation
\cite{tod} of Mach's principle, namely that there should be no
source-free contributions to the metric or that there should be no
source-free Weyl tensor for a Machian spacetime.

We, therefore, note that the vacuum is a degenerate case for
(\ref{genhsp}) and, hence, the metric (\ref{genhsp}) has no
source-free contributions. Further, the spacetime of
(\ref{genhsp}) does not possess the source-free Weyl tensor. The
spacetime of (\ref{genhsp}) is, then, also a Machian spacetime in
this sense.

%%%%%%%%%%%%%%%%%%%%%%%%%%%%%%%%%%%%%%%%%%%%%%%%%%%%%%%%%%%%%%%%%%%
\acknowledgements{ Some of the reported calculations have been
performed using the software {\tt SHEEP} and I am indebted to
Malcolm MacCallum for providing this useful software to me. I am
also grateful to Ravindra Saraykar for many helpful discussions.}
%%%%%%%%%%%%%%%%%%%%%%%%%%%%%%%%%%%%%%%%%%%%%%%%%%%%%%%%%%%%%%%%%%


\begin{thebibliography}{99}
\bibitem{mtw} See, for example, Misner C W, Thorne K S and Wheeler J A (1965)
{\it General Relativity} (New York: Freeman) \\ Wheeler J A (1964)
in {\it Relativity, Groups and Topology: Les Houches Lectures}
(New York: Gordon \& Breach)

\bibitem{sussp} See, for example, various articles in {\it Physics of the Early
Universe} (Ed. J A Peacock, A F Heavens and A T Davies)(SUSSP,
1989)

\bibitem{stdbooks} See Standard Textbooks in General Relativity, for example,
Robertson H P and Noonan T W (1969) {\it Relativity and Cosmology}
(W. B. Saunders Co., London) \\ Narlikar J V (1983) {\it
Introduction to Cosmology} (Jones and Bartlett, Boston)
\\Hawking S and Ellis, G F R (1973) {\it The large scale structure
of space-time} (Cambridge: Cambridge University Press)

\bibitem{stars} Chandrasekhar S (1958)  {\it An introduction to the study of
stellar structure} (New York: Dover) \\
Clayton D (1968) {\it Principles of stellar evolution and
nucleosynthesis} (New York: McGraw Hill)

\bibitem{mtbh} Chandrasekhar S (1983) {\it The mathematical theory of black
holes} (Oxford: Clarendon Press)

\bibitem{penroseweyl} Penrose R (1979) in {\it General Relativity -
An Einstein Centenary Survey} (Eds. S Hawking and W Israel,
Cambridge: Cambridge University Press)

\bibitem{physical} On the basis of arguments similar to those for the
spherical case in Wagh S M (2002) {\it Spherical gravitational
collapse and electromagnetic fields in radially homothetic
spacetimes} {\bf Database: gr-qc/0210020}

\bibitem{arp} Arp H (1987) in {\it Quasars, Red-shifts and
Controversies} (Berkeley: Interstellar Media) and references
therein \\ See also Burbidge G in and Hoyle F in (1988) {\it
Highlights in gravitation and cosmology} (Eds. B R Iyer, A
Kembhavi, J V Narlikar and C V Vishveshwara, Cambridge: Cambridge
University Press) \\ See also, Narlikar J V (1993) in {\it
Advances in Gravitation and Cosmology} (Eds. B R Iyer, A R
Prasanna, R K Varma and C V Vishveshwara, New Delhi: Wiley
Eastern)

\bibitem{partridge} Partridge R B (1995) {\it 3K: The Cosmic Microwave
Background Radiation} (Cambridge University Press, Cambridge)

\bibitem{cobe} Smoot G F et al. (1992) {\it Ap. J. (Lett.)}, {\bf 396}, L103.

\bibitem{0209386} Li Aigen (2002) {\it Cosmic Needles versus
Cosmic Microwave Background Radiation} {\bf Database:
astro-ph/0209386}

\bibitem{ellissciama} Ellis G F R and Sciama D (1972) In {\em
General Relativity: papers in honor of J L Synge} (Ed.
O'Raifeartaigh L, Oxford: Clarendon Press)

\bibitem{tod} Tod K P (1994) {\it Gen. Rel. Grav.} {\bf 26} 103

\end{thebibliography}
\end{document}